\documentclass[twocolumn,showpacs,preprintnumbers,amsmath,amssymb]{revtex4}
\input epsf
\begin{document}
\draft
\title{Dark propagation modes in optical lattices}

\author{M. Schiavoni, L. Sanchez-Palencia, F.-R. Carminati,
        F. Renzoni and G. Grynberg}

\address{Laboratoire Kastler Brossel, D\'epartement de Physique de l'Ecole
Normale Sup\'erieure, 24, rue Lhomond, 75231, Paris Cedex 05,
France}

\date{\today{}}

\begin{abstract}
We examine the stimulated light scattering onto the propagation modes of a
dissipative optical lattice. We show that two different pump-probe
configurations may lead to the excitation, via different mechanisms, of
the same mode. We found that in one configuration the scattering on the
propagation mode results in a resonance in the probe transmission spectrum
while in the other configuration no modification of the scattering spectrum
occurs, i.e. the mode is {\it dark}. A theoretical explanation of this
behaviour is provided.
\end{abstract}

\maketitle

\section{Introduction}

Brillouin scattering \cite{boyd,shen} is the scattering of light onto
a propagating acoustic wave. In spontaneous Brillouin scattering the 
propagating wave corresponds to thermal, or quantum-noise, fluctuations 
in the material medium. On the contrary, in stimulated Brillouin scattering 
(SBS) the density propagating wave originates from the interference pattern 
between a probe and an additional pump beam. The strong pump beam can then 
be diffracted onto the density wave in the direction of the probe, modifying 
in this way the probe transmission. The SBS-scheme permits both the excitation 
of the propagation modes of a medium as well as their detection via 
modification of the probe transmission. It is in this way possible to 
determine the phonon modes of the medium, and their respective velocity 
\cite{woolf,jap,fan}.

In this work we examine the key features of the SBS process for a nonlinear
medium consisting of atoms cooled in a dissipative optical lattice \cite{robi}.
This system offers significant advantages for the study of basic nonlinear
optical phenomena over condensed matter samples.
First, the atomic dynamics in an optical lattice is quite well understood, 
and can be precisely studied through Monte Carlo simulations. Second, the 
excitation of propagation modes in the system can be directly detected by 
imaging techniques.  Both points are essential for the present study.
We show that the {\it same} propagation mode can be excited by two different
pump-probe configurations. In one case the scattering on the propagation mode
results in a resonance in the probe transmission spectrum, while in the other 
case no modification of the spectrum occurs, i.e. the mode is in this case
{\it dark}. We describe the different excitation processes of the propagation
mode for the two different configurations examined, and identify the
mechanism of generation of the phase-mismatch between laser fields and the
material grating which inhibits the light scattering on the propagation mode.

\section{The 3D lin$\perp$lin optical lattice}

The nonlinear medium consists of $^{85}$Rb atoms cooled and trapped in a 
dissipative optical lattice. These lattices are based on the 
Sisyphus cooling mechanism \cite{dalibard89}. The periodic modulation 
of the light polarization, produced by the interference of several 
laser beams, leads to a periodic modulation of the light shifts (optical
potentials) of the different Zeeman sublevels of the ground state of the 
atom. As a result of the optical pumping between different optical 
potentials, atoms are cooled and finally trapped at the potential minima.

\begin{figure}[ht]
\begin{center}
\mbox{\epsfxsize 2.5in \epsfbox{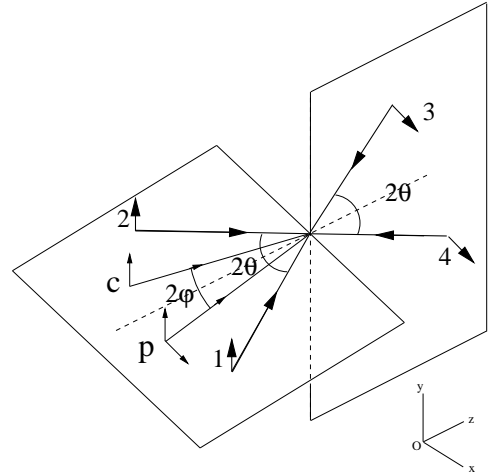}}
\end{center}
\caption{Sketch of the experimental setup. The laser fields $1 - 4$ generate
the static 3D optical potential. Two additional laser beams (c and p) are
introduced to create a moving potential modulation.}
\label{fig1}
\end{figure}

In this work we use a 3D lin$\perp$lin dissipative optical lattice \cite{robi}.
The arrangement of the laser fields is shown in Fig. \ref{fig1}: 
two $x$-polarized beams propagate in the $yOz$ plane and make an angle
$2\theta$, and two $y$-polarized beams propagate in the $xOz$ plane and 
make the same angle $2\theta$. 
The interference pattern of the four beams create an orthorhombic
potential with minima associated with pure circular
(alternatively $\sigma^{+}$ and $\sigma^{-}$) light polarization.
The lattice constants, i.e. the distance between two sites of
equal polarization are $\lambda_{x,y} = \lambda/\sin\theta$ and
$\lambda_z = \lambda/(2\cos\theta)$, with $\lambda$ the laser
field wavelength. For all the measurements presented in this work
the angle $2\theta$ between the lattice beams is kept fixed to $60^0$.

The procedure to load the atoms in the optical lattice is the standard
one used in previous experiments \cite{regis}. The rubidium atoms are
first cooled and trapped in a magneto-optical trap (MOT). Then the MOT
magnetic field and laser beams are turned off and the lattice beams are
turned on. After $10$ ms of thermalizationof the atoms in the lattice,
two additional laser fields (beams c and p of Fig. \ref{fig1}) are
introduced for the excitation of the propagation modes. They are derived
from an additional laser, with their relative detuning
$\delta=\omega_p - \omega_c$ controlled by acousto-optical modulators.
These two additional laser fields are detuned with respect to the lattice
beams of some tens of MHz, so that there is no atomic observable which can
be excited at the beat frequency. Furthermore, as they are derived from a
laser different from the one producing the lattice beams, the effect of the
unwanted beat is significantly reduced. The beams c and p cross the atomic
sample in the $xOz$ plane, and they are symmetrically displaced with respect
to the $z$ axis forming an angle $2\varphi$.

\section{Propagation modes}

\subsection{Generalities}

\begin{figure}[ht]
\begin{center}
\mbox{\epsfxsize 3.in \epsfbox{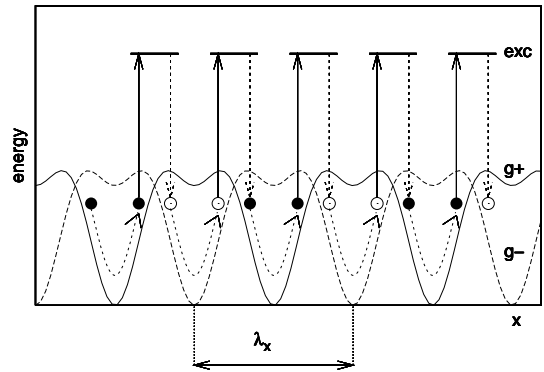}}
\end{center}
\caption{An atomic trajectory corresponding to a propagation mode in the
$x$-direction. The shown potential curves ($g_{+}$ and $g_{-}$) are the
section along $y=z=0$ of the optical potential for a $J_g=1/2\to J_e=3/2$
atomic transition and a 3D lin$\perp$lin beam configuration.}
\label{fig2}
\end{figure}

The propagation modes in dissipative optical lattices have been identified 
in Ref. \cite{brillo1} and shown to exhibit interesting nonlinear effects 
such as stochastic resonance \cite{brillo2,brillo3}.
We briefly summarize their main properties. They consist of a sequence in which 
one half oscillation in a potential well is followed by an optical pumping
process to a neighboring well, and so on (Fig. \ref{fig2}). In this way, the 
atom travels over several potential wells by regularly changing its internal
state (from $|g,+1/2\rangle$ to $|g,-1/2\rangle$ in the case of a 
$J_g=1/2\to J_e=3/2$ transition, as considered in Fig. \ref{fig2}).

The velocity $\bar{v}$ of the propagation mode is essentially determined by 
the intrawell dynamics. A straighforward calculation \cite{brillo1} shows that 
for a mode in the $x$ direction this velocity is
\begin{equation}
\bar{v} = \frac{\lambda\Omega_x}{2\pi \sin\theta}~, 
\label{vmedio}
\end{equation}
where $\Omega_x$ is the $x$ vibrational frequency at the bottom of a 
potential well.

\subsection{Excitation mechanisms}

\begin{figure}[ht]
\begin{center}
\mbox{\epsfxsize 3.in \epsfbox{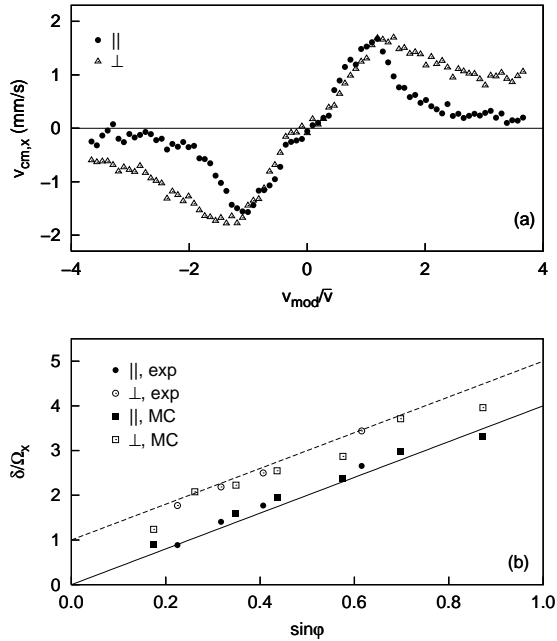}}
\end{center}
\caption{Top: experimental results for the $x$ component of the velocity of
the center-of-mass of the atomic cloud as a function of the velocity
$v_{\rm mod}$ of the moving light interference pattern. The angle between
pump and probe beams is $2\varphi=48°$. Bottom: position of the resonances
as a function of the sine of the half-angle $\varphi$ between the pump and
the probe beams. The points refer to experimental findings (expt.) and to
semiclassical Monte Carlo simulations (MC), the lines to Eqs.
(\protect\ref{eq:d_para},\ref{eq:d_perp}).}
\label{fig3}
\end{figure}

The propagation modes can be excited by adding a moving potential
modulation. We consider two different configurations for the modulation 
beams (beams c and p in Fig. \ref{fig1}). In both configurations the 
modulation beams have the same amplitude. In the first configuration, 
hereafter called the $\parallel$ configuration, both beams have $y$ linear
polarization. The light interference pattern consists of an intensity 
modulation moving along the $x$ axis with phase velocity 
\begin{equation}
v_{\rm mod} = \frac{\delta}{|\vec{\Delta k}|}= \frac{\delta}{2k
\sin\varphi}~, 
\label{vmod}
\end{equation}
where $\vec{\Delta k} = \vec{k}_p-\vec{k}_c$ is the difference between
the wavevectors of the modulation beams
($|\vec{k}_c|\cong |\vec{k}_p|\cong k\equiv 2\pi/\lambda $). This
configuration has already been considered in previous work \cite{brillo3}
and it is reexamined here for comparison with the novel excitation scheme 
introduced in the present work. This latter, denoted as $\perp$ configuration,
consists of a $y$ polarized beam (beam c of Fig. \ref{fig1}) and of a beam 
with linear polarization in the $xOz$ plane (beam p). The light interference 
pattern consists in this case of a polarization modulation moving along the 
$x$ axis with the same phase velocity $v_{\rm mod}$ (Eq.~\ref{vmod}) as in 
the case of the $\parallel$ configuration.

To determine the effective excitation of the propagation modes we monitor
the velocity of the center-of-mass (CM) of the atomic cloud as a function of
the velocity of the applied potential modulation. This is done by direct
imaging of the atomic cloud with a CCD camera. We verify that for a given
detuning $\delta$, i.e. for a given velocity $v_{\rm mod}$ of the moving
modulation, the motion of the center of mass of the atomic cloud is uniform,
and corresponding determine the CM velocity $v_{\rm cm}$. Experimental
results for the $x$ component $v_{{\rm cm},x}$ of the CM-velocity as a
function of $v_{\rm mod}$ are reported in Fig. \ref{fig3} for both the
$\parallel$ and the $\perp$ configurations. The observed resonant behaviour
of $v_{{\rm cm},x}$ with $v_{\rm mod}$ is the signature of the excitation
of propagation modes in the $x$ direction. We therefore conclude that both
pump-probe configurations lead to the excitation of a propagation mode in
the $x$ direction.

To determine the nature of the observed propagation modes, we examine the
atomic dynamics in the optical lattice with the help of semiclassical Monte
Carlo simulations \cite{epj}.
The  analysis of the numerically calculated atomic trajectories shows that 
the excited mode is the same for both configurations, and consists of a
sequence of an half oscillation in a potential well followed by an optical 
pumping into the neighbouring well, as in Fig. \ref{fig2}.

We turn now to the analysis of the excitation mechanism of the propagation
modes for the two pump-probe configuration. To this end, it is useful to
examine the dependence of the position of the resonance in the velocity of
the CM of the atomic cloud (as the ones in Fig. \ref{fig3}a) on the angle
$2\varphi$ between pump and probe beams. By taking several measurements for
different values of the angle $2\varphi$ between the modulation beams we
determine the position of the resonances as a function of the angle
$\varphi$, as reported in Fig. \ref{fig3}b. On the same plot we also
reported results of semiclassical Monte Carlo simulations, which are found
to be in very good agreement with the experimental findings. 
The results of Fig. \ref{fig3} show that the velocity $v_{\rm mod}$ of the
light intensity interference pattern ($\parallel$ configuration) required
to excite a propagation mode differs from the velocity of the polarization
grating ($\perp$ configuration) leading to the excitation of the same mode.
The condition for the velocity of the light interference pattern to excite 
a propagation mode can be obtained by imposing that the atoms following the
mode are at all times dragged by the moving potential modulation corresponding
to the light interference pattern of the beams c and p. This requires that the 
light polarization pattern moves at a velocity $v_{\rm mod}$ such that the 
lattice potential well actually occupied by the atom gets deeper as a result 
of the modulation of the optical potential. The resulting conditions on the
velocity of the moving modulation, and equivalently on the pump-probe detuning
$\delta$, can therefore be derived by examining the effect of the modulation 
on the optical potentials. 
Consider first the $\parallel$ configuration. The light interference pattern 
is a moving {\it intensity} modulation, therefore all optical potentials are
modulated {\it in phase}: at a given instant and position all  potential wells
corresponding to the different atomic ground state Zeeman sublevels get both 
deeper or shallower as a result of the modulation. Thus  to excite the
propagation mode the atoms should follow the moving intensity modulation, 
i.e. the phase velocity $v_{\rm mod}$ of the light interference pattern should 
be equal to the velocity $\bar{v}$ of the mode: 
\begin{equation}
v_{\rm mod}=\pm\bar{v}~.
\label{eq:v_para}
\end{equation}

Consider now the $\perp$ configuration. The light interference pattern is 
in this case a moving {\it polarization} modulation, with the optical 
potentials associated with opposite Zeeman sublevels (opposite quantum number 
$m$) modulated in phase opposition. It follows that in this configuration 
a modulation moving at the mode velocity does not lead to the mode excitation.
On the contrary, to excite the propagation mode it is necessary that the 
modulation moves with respect to the atoms in such a way that following the 
transfer of an atom from a lattice well of  given circular polarization 
($\sigma_{+}$ or $\sigma_{-}$) to one of opposite polarization, the modulation 
changes sign.
Quantitatively, consider the time interval $\Delta t$ in which the atom in the
propagation mode makes half an oscillation in a potential well and then is 
optically pumped into the neighbouring well. Then we simply have 
$\bar{v}\cdot\Delta t=\lambda_x/2$. In 
the same time interval the modulation polarization should be reversed, 
i.e. should change from $\sigma_{+}$ to $\sigma_{-}$ (or viceversa, 
depending on which potential well is initially occupied by the atom). 
Considering that in the time interval $\Delta t$ the atom moved of 
$\lambda_x/2$, and that in the moving modulation a maximum of polarization 
$\sigma_{-}$ is spaced of $\lambda_m/2=\pi/|\vec{\Delta k}|$ from the 
following maximum of polarization $\sigma_{+}$, we find that the light 
interference pattern should move at the velocity 
$v_{\rm mod}\cdot \Delta t = \pm( \lambda_m/2+\lambda_x/2)$.
Together with $\bar{v}\cdot \Delta t = \lambda_x/2$, we find then that 
in the $\perp$ configuration the condition for the excitation of the
propagation mode is: 
\begin{equation}
v_{\rm mod}=\pm \left(1+\frac{\sin\theta}{2\sin\varphi}\right)\bar{v}~.
\label{eq:v_perp}
\end{equation}

The conditions Eqs. (\ref{eq:v_para},\ref{eq:v_perp}) are rewritten in terms of 
the detuning $\delta$ as
\begin{mathletters}
\begin{eqnarray}
\delta & = & \pm\frac{2\sin\varphi}{\sin\theta}\Omega_x 
~~~~~~~~~ (\parallel~{\rm configuration})~, \label{eq:d_para} \\ 
\delta & = & \pm\left( 1+\frac{2\sin\varphi}{\sin\theta}\right)\Omega_x
~~~~ (\perp~{\rm configuration})~,
\label{eq:d_perp}
\end{eqnarray}
\end{mathletters}
where we used Eqs.~(\ref{vmedio},\ref{vmod}). The very good agreement (see 
Fig. \ref{fig3}) of Eqs. (\ref{eq:d_para},\ref{eq:d_perp}) with the 
experimental findings and with the results of semiclassical Monte Carlo 
simulations demonstrate the validity of our physical picture.

\subsection{Light scattering}

\begin{figure}[ht]
\begin{center}
\mbox{\epsfxsize 3.5in \epsfbox{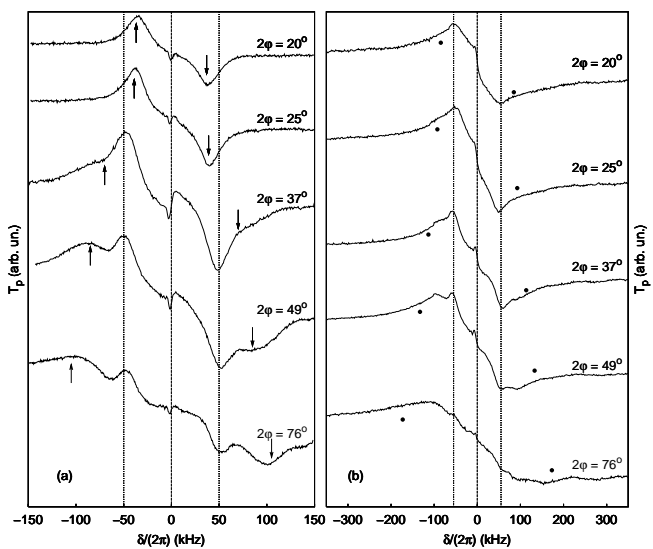}}
\end{center}
\caption{
Transmission of the probe beam as a function of the detuning between pump
and probe fields, for different values of the angle $2\varphi$ between pump
and probe beams. The lattice detuning is $\Delta=50$ MHz, the intensity per
lattice beam $I_L=5$ mW/cm$^2$. These parameters correspond to
$\Omega_x\simeq 2\pi\cdot 50$ kHz (vertical dashed lines).
The left plot corresponds to the $\parallel$ pump-probe configuration, the
right plot to the $\perp$ one.}
\label{fig4}
\end{figure}

So far we considered the effect of the light interference pattern on the
atomic sample, with the excitation of propagation modes and their detection
by direct imaging of the atomic cloud. The properties of the material medium 
can also be studied by stimulated light scattering measurements. In fact the
pump-probe interference pattern may excite a material grating onto which
the pump can be diffracted in the direction of the probe beam, modifying the
probe transmission. That is the approach followed now, with a view to compare
light scattering measurements with the previous results obtained via direct
imaging of the atomic cloud.

We decreased the amplitude of one of the modulation beams (beam p) which 
plays now the role of the probe beam, while the other beam (beam c) plays 
the role of the pump (or coupling) beam.
We measure the probe transmission as a function of the detuning $\delta$
for different angles between the pump and the probe beams, with results as 
the ones in Fig. \ref{fig4}. In the case of the $\parallel$ pump-probe
configuration, we easily identify in the probe transmission spectrum the 
Brillouin resonances (the resonances in Fig. \ref{fig4}a marked by arrows)
corresponding to stimulated light scattering on the propagation modes. We 
verified that the dependence of the position of these resonances on the 
angle $\varphi$ is in complete agreement with Eq. (\ref{eq:d_para}), which 
confirms that these resonances originate from light scattering on propagation
modes. On the contrary, in the case of the $\perp$ pump-probe configuration
no resonance is observed around the position corresponding to Eq. 
(\ref{eq:d_perp}) (these positions are marked by filled circles in Fig.
\ref{fig4}b). In other words, the propagation mode is dark in the $\perp$
configuration.

The absence of resonances in the scattering spectrum for the propagation
mode in the $\perp$ configuration can be explained by examining the
phase-mismatch between the laser and the material waves. The frequency
(energy) and phase-matching (momentum) conditions for the stimulated
scattering process read
\begin{mathletters}
\begin{eqnarray}
\omega_c = \omega_p \pm \Omega~, \label{eq:matcha} \\
\vec{k}_c = \vec{k}_p \pm \vec{q}~. \label{eq:matchb}
\end{eqnarray}
\end{mathletters}
Here $\vec{q}$ and $\Omega$ are respectively the wavevector and the
frequency of the light-induced material density grating and are
related by the phonon dispersion relation $\Omega = v_{grating}|\vec{q}|$, 
with $v_{grating}$ the phase velocity of the moving grating. The frequency
$\Omega$ has been determined previously for both $\parallel$ and $\perp$
configurations (Eqs. (\ref{eq:d_para},\ref{eq:d_perp})). As the excited mode 
is the same for both pump-probe configurations, the phase velocity of the 
material grating does not depend on the chosen configuration and is equal to
the velocity $\bar{v}$ (Eq. \ref{vmedio}). From these values for $\Omega$ and 
$v_{grating}$ we derive, through the dispersion relation, the momentum of the 
material grating:
\begin{mathletters}
\begin{eqnarray}
|\vec{q}_{\parallel}|&=& |\vec{\Delta k}|~,\label{eq:k-vecta}\\
|\vec{q}_{\perp}|&=& |\vec{\Delta k}|\left( 1 +
\frac{\sin\theta}{2\sin\varphi}\right)~.\label{eq:k-vectb}
\end{eqnarray}
\end{mathletters}

It turns out that in the $\parallel$ configuration the momentum 
$\vec{q}_{\parallel}=\pm|\Delta\vec{k}|\vec{\epsilon}_x$ of the 
material grating fulfills the phase matching condition Eq. 
(\ref{eq:matchb}), and therefore the scattering on the propagation 
mode results in a resonance line in the probe transmission spectrum. 
In contrast, in the $\perp$ configuration the momentum 
$\vec{q}_{\perp}=\pm|\vec{q}_{\perp}|\vec{\epsilon}_x$ results in a
phase-mismatch between the laser and the material waves. Thus, no resonance 
is expected in the probe transmission spectrum, in agreement with our 
experimental findings.

\begin{figure}[ht]
\begin{center}
\mbox{\epsfxsize 2.9in \epsfbox{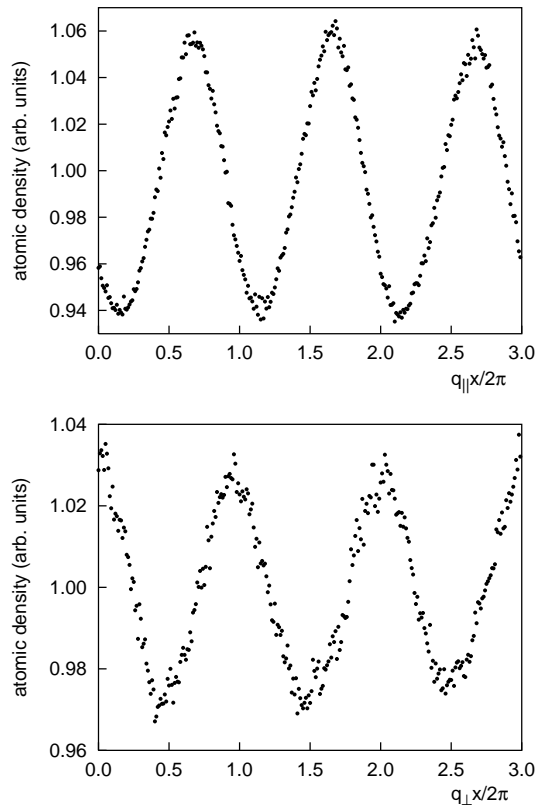}}
\end{center}
\caption{Numerical results for the atomic density as a function of $x$
for the $\parallel$ (top) and $\perp$ (bottom) pump-probe configurations.
The shown density distribution is stationary in a frame moving along $x$ at
a velocity $\bar{v}$.}
\label{fig5}
\end{figure}

The effective creation of a moving material grating has been confirmed by 
semiclassical Monte Carlo simulations. The numerical results, as the ones 
shown in Fig. \ref{fig5}, correspond to an atomic density grating moving in 
the $x$ direction with a velocity $\bar{v}$ for both parallel and 
perpendicular configurations. In the frame moving at the velocity 
$\bar{v}$ of the propagation mode we find a stationary modulation
of the atomic density with different wavevectors for the two pump-probe 
configurations and in very good agreement with Eqs. (\ref{eq:k-vecta},
\ref{eq:k-vectb}). This confirms the validity of our analysis.

\section{Conclusions}

In summary, in this work we examined the stimulated light scattering onto
the propagation modes of a dissipative optical lattice. Two different 
pump-probe configurations have been analyzed: in one the interference 
pattern is a modulation of the light intensity, while in the other one 
the pump and probe fields give rise to a modulation of the light 
polarization. First, we have shown that the {\it same} propagation mode 
is excited in the two cases, and described the two different mechanisms of 
excitation.
Then we analyzed the light scattering on the propagation mode. Although 
the mode excited in the two pump-probe configurations is the same, we found
that the probe transmission spectrum is completely different for the two 
cases. In fact, only in one configuration the mode results in a resonance
in the probe transmission spectrum. For the other configuration, no trace
of the mode excitation is found in the probe transmission spectrum. This
behaviour was explained in terms of phase-mismatch between the laser fields
and the propagating wave.

Light scattering is a powerful technique for the study of a large variety of
material media. Particularly, in optical lattices it has allowed the study of 
local (intrawell) as well as delocalized (inter wells) dynamics. However there
is not a one-to-one correspondence between the light scattering spectrum and 
the atomic dynamics as shown in this work where we observed and described
dark propagation modes. 

This work was supported by R\'egion Ile de France under contract E.1220.
Laboratoire Kastler Brossel is an "unit\'e mixte de recherche de
l'Ecole Normale Sup\'erieure et de l'Universit\'e Pierre et Marie
Curie associ\'ee au Centre National de la Recherche Scientifique
(CNRS)".

\end{document}